\documentclass{iaus}
\usepackage{graphicx}

\title[M67 in the VO] 
{Stellar Population Challenge: analysis of M67 with the VO}

\author[Cervi\~no \& Guti\'errez]   
{M. Cervi\~no$^{1,3}$,  
 R. Guiti\'errez$^{2,3}$  \and E. Solano$^{2,3}$}

\affiliation{$^1$Instituto de Astrof\'\i sica de Andaluc\'\i a (IAA-CSIC), Camino Bajo de Hu\'etor 50, 18008 Spain\break email: mcs@iaa.es\\[\affilskip]
$^2$Laboratorio de Astrof\'\i sica Espacial y F\'\i sica Fundamental (LAEFF-INTA), Apdo 50727, Madrid 28080\break email: raul@laeff.inta.es, esm@laeff.inta.es\\[\affilskip]
$^3$Spanish Virtual Observatory (SVO)  {\tt http://svo.laeff.inta.es}
}

\pubyear{2007}
\volume{241}  
\pagerange{119--126}
\date{?? and in revised form ??}
\jname{Stellar Populations as Building Blocks of Galaxies}
\editors{A. Vazdekis \& R. Peletier, eds.}
\begin{document}

\maketitle

\begin{abstract}
In this poster we present the analysis of the CMD of M67 (proposed in the Stellar Population Challenge) performed with VO applications. We found 
that, although the VO environment is still not ready to perform a complete analysis, its use provides highly useful additional information for the 
analysis. Thanks to the current VO framework, we are able to identify stars in the provided CMD that are not suitable for isochrone fitting. Additionally, 
we can complete our knowledge of this cluster extending the analysis to IR colors, which were not provided in the original data but that are available 
thanks to the VO. On the negative side, we find it difficult to access theoretical data from VO applications, so, currently, it is not possible to perform completely the analysis of the cluster inside the VO framework. However it is expected that the situation will improve in a near future.

\keywords{galaxies: star clusters}
\end{abstract}

\firstsection 

\section{M67 in the VO}

\begin{figure}
 \includegraphics[height=2.3in]{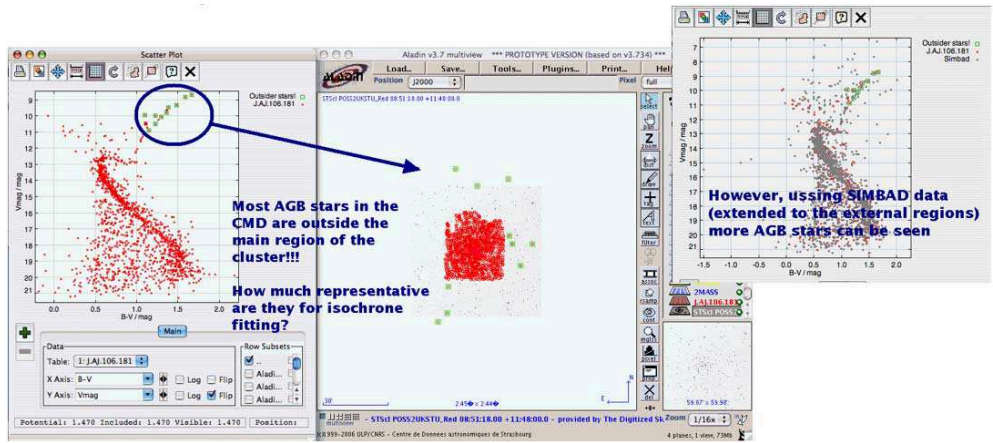}
  \caption{Screenshoot of the analysis process of M67. In the left panel the position of stars is shown
simultaneously in the field and in the CMD diagram.
The right panel shows the CMD ussing additional SIMBAD data.}
  \label{fig1}
\end{figure}

Following the {\it Stellar Population Challenge} proposed in this meeting, we have take the \cite{Mont} data of the CMD of M67 (in B, V and R bands) and we have analyzed it with ALADIN (\cite{Aladin})\footnote{{\tt http://aladin.u-strasbg.fr/aladin.gml}} and TOPCAT\footnote{{\tt http://www.star.bristol.ac.uk/~mbt/topcat/}}. Taking advantage of the Virtual Observatory, we have included in the analysis additional stars from SIMBAD and we have obtained the photometry of such stars in Infrared bands from 2MASS. This process is illustrated by the screenshoot showed in Fig. \ref{fig1}. This first step can be performed quickly and smoothly by the use of current VO tools.

\begin{figure}
 \includegraphics[height=3in]{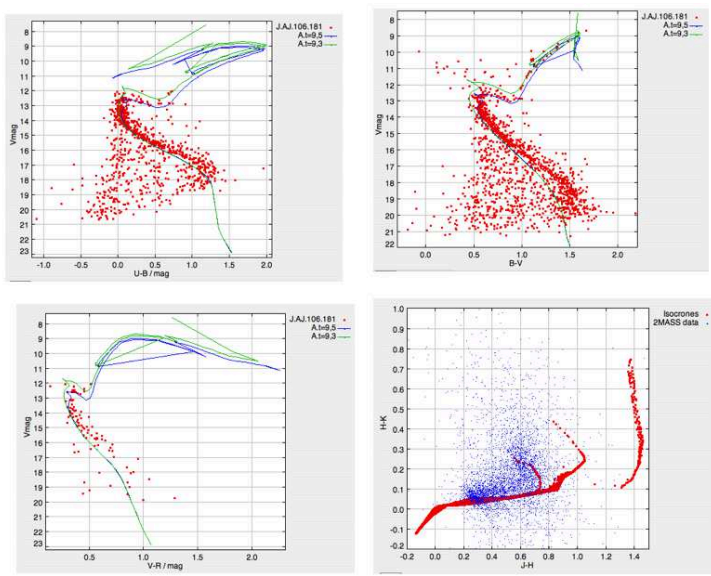}
  \caption{Different CMD fits performed by eye.}
  \label{fig2}
\end{figure}

In a second step, we have obtained the isochrones by \cite{Cioni06a,Cioni06b} from the Padova Web server\footnote{{\tt http://pleiadi.pd.astro.it/}} since it is not trivial to access them directly form the VO. Fig. \ref{fig2} shows the best fitting isochrones ($\log t= 9.3, 9.5$). This last fitting has been performed by eye.

From this exercise, we found that VO tools make a very good job in the area of obtain and interoperating different observational datasets. However, they are still not ready to perform a complete analysis including theoretical isochrones. It is due to the lack of an standard access to theoretical data, and, consequently, the absence of general fitting tools that combine theoretical and observational data.  However, there are some efforts to overcome this difficulty, although focused in particular astrophysical problems different than CMD fitting. We hope that the situation will be improved in a near future.

\begin{acknowledgments}
We acknowledge Mark Taylor by the development of TOPCAT and CDS for the development of Aladin, SIMBAD and VizieR. 
This work was supported by the Spanish project PNAYA2005-24102-E. MC is supported by a Ram\'on y Cajal fellowship.
\end{acknowledgments}

\end{document}